\documentclass[a4paper,fleqn,usenatbib,useAMS]{mnras}

\usepackage{float}
\usepackage{graphicx}
\usepackage{times}
\usepackage{amstext}
\usepackage{amsmath}
\usepackage{amssymb}	% Extra maths symbols
\usepackage{natbib}
\usepackage{url}
\usepackage{tabularx}
\usepackage{mathtools}
\usepackage{mathrsfs}
\usepackage{graphics}
\usepackage{multirow}
\usepackage{ar}

\newcommand{\kms}{\,km\,s$^{-1}$} % kilometres per second
\newcommand{\ntwoh}{N$_{2}$H$^{+}$}

\newcommand{\tone}{$J=1\rightarrow0$}

\newcommand{\vel}{km\,s$^{-1}$\,pc$^{-1}$}
\newcommand{\scouse}{{\sc scouse}}
\newcommand{\solar}{M$_{\odot}$}

\usepackage[T1]{fontenc}
\usepackage{ae,aecompl}

\usepackage{mathptmx}
\usepackage{txfonts}

\title[Seeding the Galactic Centre gas stream]{Seeding the Galactic Centre gas stream: gravitational instabilities set the initial conditions for the formation of protocluster clouds}

\author[J. D. Henshaw et al.]{J. D. Henshaw$^{1}$\thanks{Contact e-mail: j.d.henshaw@ljmu.ac.uk}, S. N. Longmore$^{1}$, and J. M. D. Kruijssen$^{2,3}$, \\
$^{1}$ Astrophysics Research Institute, Liverpool John Moores University, Liverpool, L3 5RF, UK\\
$^{2}$ Astronomisches Rechen-Institut, Zentrum f{\"u}r Astronomie der Universit{\"a}t Heidelberg, M{\"o}nchhofstra{\ss}e 12-14, D-69120 Heidelberg, Germany\\
$^{3}$Max-Planck Institut f\"{u}r Astronomie, K\"{o}nigstuhl 17, D-69117 Heidelberg, Germany }

\date{Last updated XXXX; in original form XXXX}

\pubyear{2016}

\begin{document}
\label{firstpage}
\pagerange{\pageref{firstpage}--\pageref{lastpage}}
\maketitle

\begin{abstract}
Star formation within the Central Molecular Zone (CMZ) may be intimately linked to the orbital dynamics of the gas. Recent models suggest that star formation within the dust ridge molecular clouds (from G0.253+0.016 to Sgr B2) follows an evolutionary time sequence, triggered by tidal compression during their preceding pericentre passage. Given that these clouds are the most likely precursors to a generation of massive stars and extreme star clusters, this scenario would have profound implications for constraining the time-evolution of star formation. In this Letter, we search for the initial conditions of the protocluster clouds, focusing on the kinematics of gas situated upstream from pericentre. We observe a highly-regular corrugated velocity field in $\{l,\,v_{\rm LSR}\}$ space, with amplitude and wavelength $A=3.7\,\pm\,0.1$\,\kms \ and $\lambda_{\rm vel, i}=22.5\,\pm\,0.1$\,pc, respectively. The extremes in velocity correlate with a series of massive ($\sim10^{4}$\,\solar) and compact ($R_{\rm eq}\sim2$\,pc), quasi-regularly spaced ($\sim8$\,pc), molecular clouds. The corrugation wavelength and cloud separation closely agree with the predicted Toomre ($\sim17$\,pc) and Jeans ($\sim6$\,pc) lengths, respectively. We conclude that gravitational instabilities are driving the condensation of molecular clouds within the Galactic Centre gas stream. Furthermore, we speculate these seeds are the historical analogue of the dust-ridge molecular clouds, representing the initial conditions of star and cluster formation in the CMZ.

\end{abstract}

\begin{keywords}
stars: formation -- ISM: clouds -- ISM: kinematics and dynamics -- ISM: structure -- Galaxy: centre -- Galaxy: kinematics and dynamics
\end{keywords}

%%%%%%%%%%%%%%%%%%%%%%%%%%%%%%%%%%%%%%%%%%%%%%%%%%

%%%%%%%%%%%%%%%%% BODY OF PAPER %%%%%%%%%%%%%%%%%%

\section{Introduction}

The ultimate goal of star formation studies is an end-to-end, temporal understanding of stellar mass assembly as a function of environment. The Central Molecular Zone (CMZ, i.e.~the central few 100 pc of the Milky Way) hosts some of the most extreme products of this process within the Galaxy. These include massive star clusters, such as the Arches and Quintuplet \citep{portegies_2010}, and massive ($10^{5}-10^{6}$\,\solar) protocluster clouds, such as Sgr B2, which displays prominent star formation activity (evidenced through, ultra-compact, compact, and large H{\sc ii} regions, and maser activity; \citealp{gaume_1995, mehringer_1997, mcgrath_2004, de-pree_2014}). The CMZ contains a significant fraction ($\sim80$ per cent) of the Milky Way's dense ($\gtrsim10^{3}$\,cm$^{-3}$) molecular gas \citep{molinari_2014}. Somewhat counterintuitively however, the present-day star formation rate of the CMZ ($\sim0.05$\,\solar\,yr$^{-1}$; \citealp{crocker_2012, longmore_2013a}), is much lower than expected, if this dense gas were forming stars on a similar timescale to that found in the Galactic disc.

Recent theoretical developments suggest that the CMZ currently resides in a period of quiescence \citep{kruijssen_2014b,krumholz_2015,krumholz_2016}. In these models, highly super-virial gas is driven inwards from large-scales by angular momentum transport induced by acoustic instabilities. The turnover in the Galactic rotation curve from approximately flat to approximately solid body, and the subsequent reduction in shear and radial transport, results in the build up of gas within a radial annulus at a galactocentric radius of $\sim100$\,pc. A sleeping giant, the CMZ gas stream becomes episodically gravitationally-unstable on time-scales of 10-20\,Myr, resulting in rapid star formation.

\begin{figure*}
\begin{center}
\includegraphics[trim = 0mm 0mm 0mm 0mm, clip, width = 0.95\textwidth]{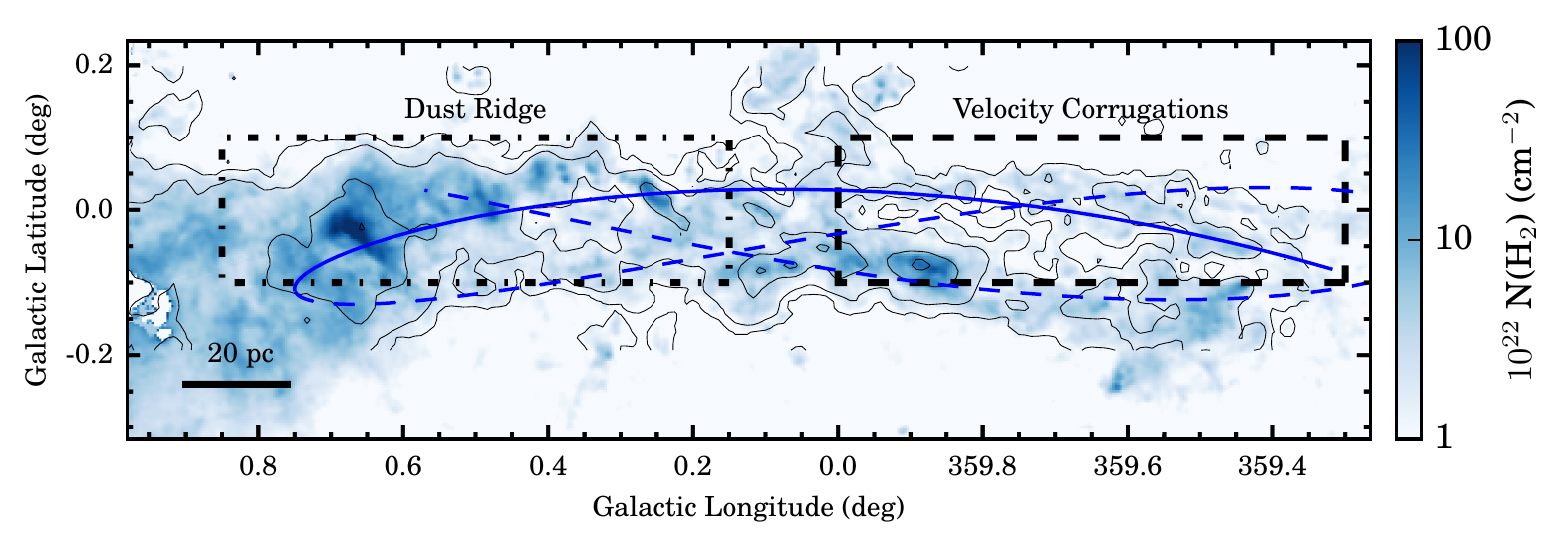}
\end{center}
\vspace{-0.5cm}
\caption{The \emph{Herschel}-derived molecular hydrogen column density map of the CMZ (Battersby et al.~in preparation). The black contours represent the integrated emission from \ntwoh \ \tone. Contour levels represent 1, 10, 50, and 90 per cent of peak emission. The orbital solution of \citet{kruijssen_2015} is highlighted by the blue line. The portion relevant to this study is indicated by the solid line. The black dashed and black dot-dashed rectangle reflect the region of interest to the present study and the dust ridge, respectively. The scale bar  assumes a distance of 8.3~kpc \citep{reid_2014}. }
\label{Figure:CMZ_map}
\end{figure*}

An important ingredient for the initiation of star formation within the inner gas stream relates to the orbital dynamics of the gas. Observational studies of the dust ridge molecular clouds (\citealp{lis_2001}; see the dot-dashed box in Fig.~\ref{Figure:CMZ_map}), which indicate a sequential increase in star formation activity as a function of increasing Galactic longitude, led \citet{longmore_2013b} to postulate that these protocluster clouds may share a common timeline, and hence represent an evolutionary sequence. The sequence commences with G0.253+0.016, which shows observational signatures of, but overall very little star formation activity (e.g. \citealp{mills_2015}), through clouds d, e, and f (cloud e shows class {\sc ii} methanol and water maser activity; \citealp{forster_1999,valdettaro_2001, caswell_2010}), towards Sgr B2, which displays prominent star formation activity (see above). The coherency of the dust ridge gas stream is supported by the comprehensive decomposition of the complex molecular gas kinematics performed by \citet{henshaw_2016}. In this scenario, star formation is initiated at a common zero point, triggered as the gas clouds are tidally compressed during their pericentre passage at $\sim60$ pc from the bottom of the Galactic gravitational potential \citep{kruijssen_2015}. Dynamical modeling of the gas stream performed by \citet{kruijssen_2015}, predicts that G0.253+0.016 and Sgr B2 are separated by $0.43^{+0.22}_{-0.08}$ Myr, which corresponds to a single free-fall time ($t_{\rm ff}\sim0.34$\,Myr). In the context of the proposed evolutionary timeline, this result suggests that once star formation has been initiated, it happens rapidly.

The question is whether it is reasonable to assume that the clouds have similar initial conditions, such that the tidal compression at pericentre causes them to evolve and form stars on similar time-scales. To constrain the initial conditions of the dust ridge clouds, we must search upstream. In this Letter, we investigate the characteristic properties and origins of the fluctuating velocity field of the upstream gas, first identified by \citet{henshaw_2016}. We highlight the correlation between the extremes of the velocity field and peaks in the density field, which indicates the convergence of gas flows onto the seeds of molecular clouds. Demonstrating the similarity between the observed corrugation wavelength and the Toomre length, and between the observed cloud spacing and the predicted Jeans length, we conclude that gravitational instabilities drive the formation of molecular clouds within the Galactic Centre (hereafter, GC) gas stream.

\section{Data}

This paper makes use of data from the Mopra CMZ survey, first published by \citet{jones_2012}. We focus on the \tone \ transition of \ntwoh \ (the rest frequency of the main ${\rm J,\,F_{1},\,F}~=~1,\,2,\,3 \rightarrow 0,\,1,\,2$ hyperfine component is $\sim 93.174$\,GHz), the emission from which is both bright and contiguous over the region of interest \citep{henshaw_2016}. The presented data cubes have an effective spatial resolution of $\sim60$\,arcsec (corresponding to $\sim2.4$\,pc), a pixel size of 30\,arcsec\,$\times$\,30\,arcsec ($1.2\,{\rm pc}\times1.2\,{\rm pc}$), a spectral resolution of 2\,\kms, and the typical rms noise level is $\sim0.01-0.02$\,K. Figure~\ref{Figure:CMZ_map} is a map of the inner CMZ. The colour-scale represents the \emph{Herschel}-derived column density (Battersby et al.~in preparation). The present investigation focuses on the region $-0\fdg7<l<0\fdg0$, $-0\fdg1<b<0\fdg1$, which is indicated by the black dashed rectangle. The black contours highlight integrated emission of \ntwoh. The data enclosed within the black dashed rectangle were analysed using the Semi-automated multi-COmponent Universal Spectral-line fitting Engine (\scouse\footnote{\scouse \ is available at \url{https://github.com/jdhenshaw/SCOUSE}}; \citealp{henshaw_2016}).

\section{Results}\label{Section:results}

\begin{figure}
\begin{center}
\includegraphics[trim = 0mm 20mm 0mm 30mm, clip, width = 0.48\textwidth]{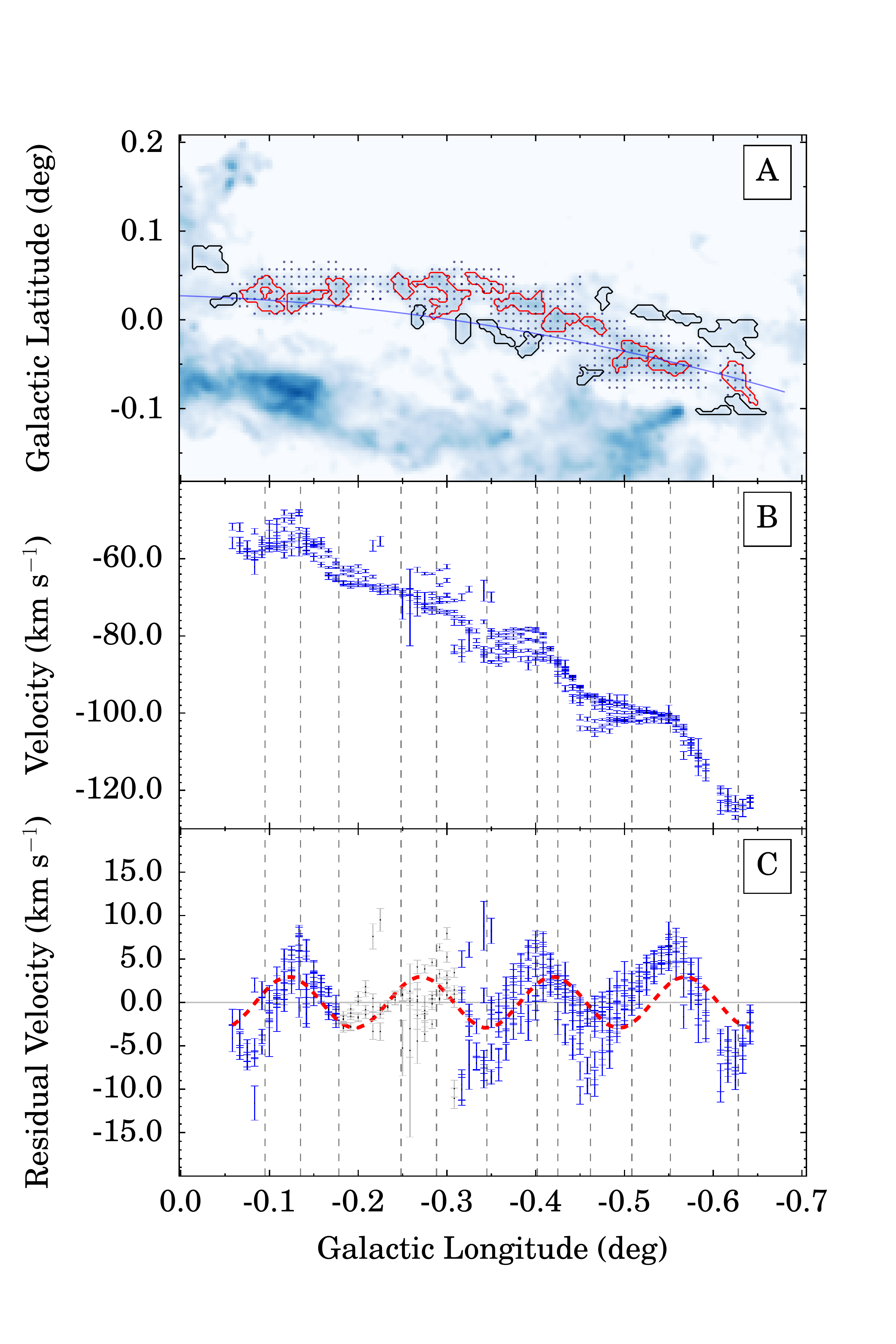}
\end{center}
\caption{Galactic longitude versus: (A) Galactic latitude. A column density map of the CMZ focusing on the region of interest (Battersby et al. in preparation). The colour scale and blue solid line are equivalent to those in Fig.~\ref{Figure:CMZ_map}. The contours refer to dendrogram-identified structures. The red contours are those spatially associated with the corrugated velocity field; (B) $v_{\rm LSR}$. The data points refer to the centroid velocities of spectral components extracted using \scouse \ (blue data points in panel A); (C) residual velocity after removal of a linear velocity gradient (Equation~\ref{Eq:lin_grad} and \S~\ref{Section:results}). The red dashed line represents the best-fitting solution to a sinusoidal model used to describe the blue data points. Grey data points indicate those that are excluded from the fitting process (see \S~\ref{Section:results}). }
\label{Figure:wiggles}
\end{figure}

Panel A of Fig.~\ref{Figure:wiggles} provides a close-up view of the \emph{Herschel}-derived column density map (Fig.~\ref{Figure:CMZ_map}). It focuses on the location of the upstream velocity field (depicted by the blue dots). To identify the main structures we use dendrograms \citep{rosolowski_2008}. Each contour in panel A represents a leaf, the highest level in the dendrogram hierarchy.\footnote{The following parameters are used in computing the dendrogram: ${\tt min\_value}=~1.5\times10^{22}$\,cm$^{-2}$; ${\tt min\_delta}=~0.5\times10^{22}$\,cm$^{-2}$; ${\tt min\_npix}=~20$. Although the hierarchical structure is sensitive to variation in the input parameters, the leaves are robust. Only two out of the twelve leaves fragment when reducing ${\tt min\_npix}$ by a factor of 2.} Red contours depict structures that are spatially coincident with the velocity pattern. The column density map of Battersby et al. (in preparation) is used to derive physical properties of each leaf, including the peak column density ($N$), total mass ($M$), equivalent number density and corresponding equivalent free-fall time ($n_{\rm eq}$ and $t_{\rm ff, eq}$).\footnote{The \emph{equivalent} number density, $n_{\rm eq}$ and free-fall time, $t_{\rm ff, eq}$, refer to that of a spherical cloud of equivalent radius, $R_{\rm eq}=(N_{\rm pix}A_{\rm pix}/\pi)^{1/2}$, where $N_{\rm pix}$ and $A_{\rm pix}$ refer to the total number of leaf pixels and the area of a single pixel, respectively. } These properties are presented in Table~\ref{Table:clouds}. 

Panel B of Fig.~\ref{Figure:wiggles} showcases the velocity pattern first identified by \citet{henshaw_2016}, which appears as a corrugation in $\{l,\,v_{\rm LSR}\}$ space. Each data point refers to the location ($l$) and centroid velocity ($v_{\rm 0}$) of a Gaussian component extracted using \scouse. We corroborated these values using the hyperfine structure (HFS) fitting functionality available in {\sc class}.\footnote{\url{http://www.iram.fr/IRAMFR/GILDAS}} This method accounts for the effects of opacity on the line-profiles (whereas Gaussian fitting does not), leading to a more accurate measure of the velocity dispersion and line centroid where the opacity is high. We find that the velocity dispersions derived using \scouse \ are systematically larger than those obtained using {\sc class} ($\langle\sigma^{\rm {\scouse}}/\sigma^{\rm {\sc class}}\rangle\sim1.7$). However, the derived centroid velocities are very similar, $\langle|v^{\rm {\scouse}}_{0}-v^{\rm {\sc class}}_{0}|\rangle\sim1.5$\,\kms. Both the \scouse \ and {\sc class} derived kinematic properties are included in Table~\ref{Table:clouds}.

To analyse the oscillatory pattern, we first remove a linear gradient of the form
\begin{equation}
v_{\rm grad}=v_{c}+\nabla v_{l}\Delta l+\nabla v_{b}\Delta b
\label{Eq:lin_grad}
\end{equation}
where $v_{c}$ is a constant, $\Delta l$ and $\Delta b$ are the offset Galactic longitude and latitude, and $\nabla v_{l}$ and $\nabla v_{b}$ are the magnitudes of the velocity gradients in the $l$ and $b$ directions, respectively. In practise we have inserted a break point at $l=-0\fdg33$, where the coherency of the velocity pattern diminishes, and fit two gradients, one above and one below this point. The least-squares minimization routine {\sc mpfit} \citep{markwardt_2009} is used to find the best-fitting solution to Equation~\ref{Eq:lin_grad} in both instances. We find that the overall magnitude of the velocity gradient at longitudes less than and greater than the break point is, $0.97\,\pm\,0.01$\,\vel \ and $0.96\,\pm\,0.02$\,\vel, respectively. The gradients are directed $85\fdg9\,\pm\,0\fdg5$ and $139\fdg5\,\pm\,1\fdg5$ east of Galactic north, respectively.

\begin{figure}
\begin{center}
\includegraphics[trim = 0mm 215mm 0mm 30mm, clip, width = 0.48\textwidth]{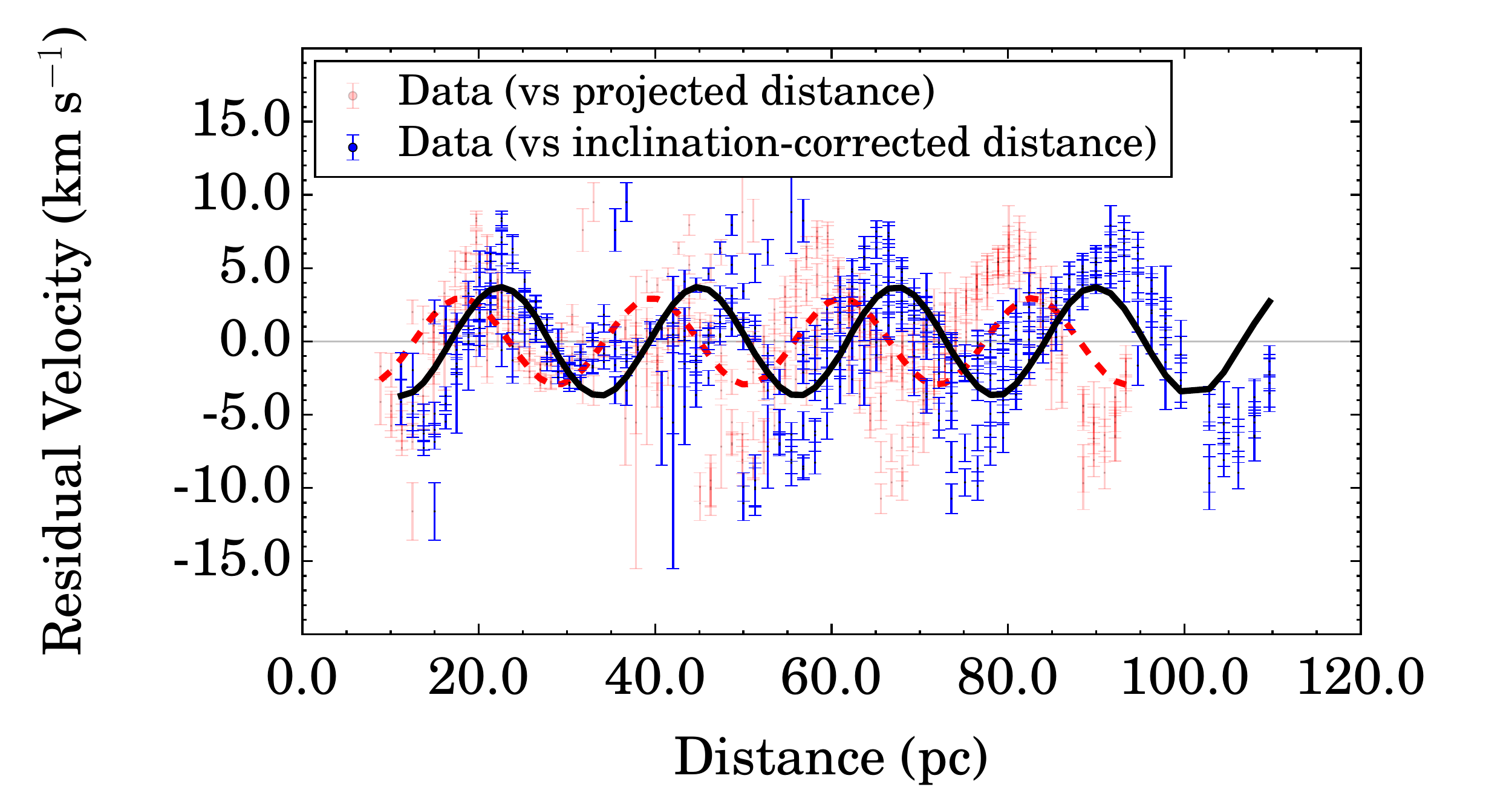}
\end{center}
\caption{Residual velocity versus: (red) projected distance (from $\{l,\,b\}\approx~\{0\fdg0,\,0\fdg0\}$); (blue) inclination-correction distance along the orbit after adopting the three-dimensional geometry of \citet{kruijssen_2015}. The corresponding best-fit model solutions are shown as dashed red and solid black lines, respectively. }
\label{Figure:wiggles_comparison}
\end{figure}

\begin{table*}
	\caption{Dendrogram leaves: kinematic and physical properties. Columns: ($a$) Leaf position; (${b}$) Equivalent radius; $R_{\rm eq}=(N_{\rm pix}A_{\rm pix}/\pi)^{1/2}$ (see text for details); (${c}$) \scouse-derived centroid velocity and velocity dispersion (see \S~\ref{Section:results})); (${d}$) {\sc class}-derived centroid velocity, velocity dispersion, and optical depth (see \S~\ref{Section:results}). (${e}$) Residual velocity (see \S~\ref{Section:results}); (${f}$) Peak column density. (${g}$) Leaf mass; (${h}$) Equivalent number density; $n_{\rm eq}=3M/4\pi R^{3}_{\rm eq}\mu m_{\rm H}$, where $\mu=2.8$ and $m_{\rm H}$ is the mass of a hydrogen atom; (${i}$) Equivalent free-fall time; $t_{\rm ff, eq}=(3\pi/32G\mu m_{\rm H}n_{\rm eq})^{1/2}$. 
	} \vspace{0.2cm}
	
	\centering  
	\tabcolsep=0.1cm \small{
	\begin{tabular}{  c  c  c  c  c  c  c  c  c  c  c  c  c  c  }
	\hline
	ID & 
	$l^{a}$ & 
	$b^{a}$ & 
	$R_{\rm eq}^{b}$ & 
	$v^{c}_{0}$ & 
	$v^{d}_{0}$ & 
        $\sigma^{c}_{v}$ &
	$\sigma^{d}_{v}$ &
    $\tau^{d}$ &
    $v_{\rm resid}^{e}$ &
    $N^{f}$ &
	$M^{g}$ &
	$n_{\rm eq}^{h}$ &
    $t_{\rm ff, eq}^{i}$
	\\ [0.5ex]
    
    & 
	& 
	& 
	& 
	& 
	& 
	&
    &
	&
    &
    $\times10^{22}$ &
	$\times10^{3}$ &
	$\times10^{3}$ &
    $\times10^{6}$ 
	\\ [0.5ex]
    
        & 
	(deg) & 
	(deg) & 
	(pc) & 
	(km\,s$^{-1}$) & 
	(km\,s$^{-1}$) &
	(km\,s$^{-1}$) &
        (km\,s$^{-1}$) &
        &
        (km\,s$^{-1}$) &
        (cm$^{-2}$) &
	(\solar) &
	(cm$^{-3}$) &
    (yr)
	\\ [0.5ex]
	
	\hline 

 0 & -0.625 & -0.068 &  2.3 & -125.19 ( 0.42 ) & -126.05 ( 0.36 ) &  8.67 ( 0.42 ) &  3.12 ( 0.83 ) &  2.67 ( 2.02 ) & -7.23 ( 1.03 ) &  3.5 &  8.5 &  2.4 & 0.63 \\ [0.5ex]
 1 & -0.550 & -0.051 &  1.9 & -100.65 ( 0.12 ) & -101.77 ( 0.12 ) &  9.61 ( 0.12 ) &  5.91 ( 0.20 ) &  2.58 ( 0.30 ) &  6.60 ( 0.87 ) &  7.1 & 13.7 &  6.8 & 0.37 \\ [0.5ex]
 2 & -0.508 & -0.035 &  2.2 & -99.82 ( 0.29 ) & -101.35 ( 0.14 ) &  9.10 ( 0.29 ) &  6.01 ( 0.30 ) &  1.98 ( 0.41 ) &  1.40 ( 0.87 ) &  7.0 & 17.9 &  5.6 & 0.41 \\ [0.5ex]
 3 & -0.458 & -0.001 &  1.5 & -95.45 ( 0.13 ) & -98.23 ( 0.24 ) &  8.86 ( 0.13 ) &  4.89 ( 0.82 ) &  1.61 ( 1.26 ) & -1.61 ( 0.78 ) &  5.9 &  6.3 &  6.9 & 0.37 \\ [0.5ex]
 4 & -0.425 &  0.007 &  2.2 & -86.72 ( 0.18 ) & -87.63 ( 0.15 ) & 10.32 ( 0.18 ) &  8.52 ( 0.12 ) &  0.10 ( 0.09 ) &  2.36 ( 0.76 ) &  6.0 & 13.0 &  4.4 & 0.46 \\ [0.5ex]
 5 & -0.400 &  0.015 &  2.4 & -77.98 ( 0.11 ) & -79.08 ( 0.11 ) &  8.89 ( 0.11 ) &  5.04 ( 0.23 ) &  3.37 ( 0.50 ) &  7.50 ( 0.45 ) &  5.6 & 15.3 &  3.9 & 0.50 \\ [0.5ex]
 6 & -0.342 &  0.040 &  1.8 & -84.46 ( 0.28 ) & -86.48 ( 0.31 ) &  8.75 ( 0.28 ) &  7.87 ( 0.41 ) &  0.10 ( 0.07 ) & -7.45 ( 0.61 ) &  2.7 &  5.2 &  3.2 & 0.55 \\ [0.5ex]
 7 & -0.292 &  0.049 &  2.9 & -73.19 ( 0.14 ) & -74.38 ( 0.20 ) &  8.00 ( 0.14 ) &  4.67 ( 0.61 ) &  2.52 ( 1.09 ) &  0.41 ( 0.49 ) &  5.3 & 18.8 &  2.6 & 0.60 \\ [0.5ex]
 8 & -0.250 &  0.040 &  1.7 & -68.16 ( 0.50 ) & -69.35 ( 0.43 ) &  5.67 ( 0.50 ) &  2.46 ( 0.21 ) &  3.13 ( 0.44 ) &  0.82 ( 0.73 ) &  4.2 &  6.4 &  4.6 & 0.45 \\ [0.5ex]
 9 & -0.175 &  0.032 &  1.9 & -62.37 ( 0.15 ) & -65.37 ( 0.07 ) &  7.96 ( 0.15 ) &  5.92 ( 0.14 ) &  0.10 ( 0.00 ) & -1.00 ( 0.36 ) &  4.4 &  8.3 &  4.2 & 0.47 \\ [0.5ex]
10 & -0.133 &  0.024 &  2.0 & -48.34 ( 0.39 ) & -50.06 ( 0.45 ) & 11.91 ( 0.39 ) &  8.83 ( 0.54 ) &  1.33 ( 0.53 ) &  8.40 ( 0.63 ) &  6.9 & 10.8 &  4.4 & 0.46 \\ [0.5ex]
11 & -0.092 &  0.040 &  2.7 & -58.46 ( 0.52 ) & -58.96 ( 0.39 ) &  8.46 ( 0.52 ) &  6.30 ( 0.61 ) &  0.10 ( 0.24 ) & -3.71 ( 0.74 ) &  4.6 & 15.5 &  2.6 & 0.60 \\ [0.5ex]

	\hline 
\end{tabular}
%	\vspace{0.2cm}
%
%\begin{minipage}{1.0\textwidth}\footnotesize{
%$^{a}$ Leaf position. This refers to the location of peak column density of each leaf. \\
%$^{b}$ Equivalent radius; $R_{\rm eq}=(N_{\rm pix}A_{\rm pix}/\pi)^{1/2}$, where $N_{\rm pix}$ and $A_{\rm pix}$ refer to the total number of pixels the area of a single pixel, respectively. \\%
%$^{c}$ \scouse-derived Gaussian centroid velocity and velocity dispersion (see \S~\ref{Section:results})). \\
%$^{d}$ {\sc class}-derived HFS centroid velocity, velocity dispersion, and optical depth (see \S~\ref{Section:results}). \\
%$^{e}$ Residual velocity after subtraction of a linear velocity gradient (see \S~\ref{Section:results}). \\
%$^{f}$ Peak \emph{Herschel}-derived column density. \\
%$^{g}$ Leaf mass. \\
%$^{h}$ Equivalent number density; $n_{\rm eq}=[M/(\frac{4}{3}\pi R^{3}_{\rm eq}\mu m_{\rm H}]$, where $\mu=2.8$ and $m_{\rm H}$ is the mass of a hydrogen atom. \\
%$^{i}$ Equivalent free-fall time; $t_{\rm ff, eq}=[3\pi/(32G\mu m_{\rm H}n_{\rm eq})^{1/2}]$. \\
%}
%\end{minipage}
}
\label{Table:clouds}
\end{table*}

Panel C of Fig.~\ref{Figure:wiggles} shows the residual velocity, $v_{\rm resid}$, after subtracting the contribution to the velocity from the underlying linear gradient ($v_{\rm resid}=v_{0}-v_{\rm grad}$). This highlights the regularity of the corrugated velocity pattern.\footnote{The velocity gradients parallel to the stream give the appearance of a corrugation in $\{l,\,v_{\rm LSR}\}$ space. However, the velocities also vary \emph{perpendicular} to the stream. This leads to the broadening in the velocity pattern evident in Fig.~\ref{Figure:wiggles} panel C.} We model the velocity field using a sinusoidal function of the form
\begin{equation}
v_{\rm resid}=A{\rm sin}(2\pi l/\lambda_{\rm vel})
\label{Eq:sin}
\end{equation}
where $A$ and $\lambda_{\rm vel}$ represent the amplitude and the wavelength of the corrugation, respectively. To fit the velocity field we mask the data between $-0\fdg31<l<-0\fdg18$. This coincides with a decrease in the ratio between peak and median stream surface density of $\sim$ a factor of 2, and is where the coherency of the signal diminishes (grey data points in Fig.~\ref{Figure:wiggles}). The resulting best-fitting solution to Equation~\ref{Eq:sin} has $A=3.0\pm0.1$\,\kms \ and $\lambda_{\rm vel}=21.6\pm0.1$\,pc, with $\chi^{2}_{\rm red}\approx3.3$ (dashed red line; Fig.~\ref{Figure:wiggles}). 

Although a reasonable fit is achieved, it is notable that the period of oscillation appears to change over the extent of the gas stream. This is likely caused by inclination and orbital curvature. Since the GC gas stream is seen in projection, the actual wavelength of the corrugation may differ from that observed, and differentially so if the orbit is curved. Accounting for inclination necessitates knowledge of the three-dimensional structure of the CMZ. For this, we adopt the orbital solution of \citet{kruijssen_2015}. We relate the cumulative distance along the orbit to Galactic longitude and estimate how inclination affects the corrugation wavelength. Modeling these data using a sinusoidal function (Equation~\ref{Eq:sin}), we find that the best-fitting solution has $A=3.7\pm0.1$\,\kms \ and $\lambda_{\rm vel, i}=22.5\pm0.1$\,pc, with $\chi^{2}_{\rm red}\approx2.7$. The inclination-corrected data and best-fitting solution are shown in Fig.~\ref{Figure:wiggles_comparison}.

Also notable in Fig.~\ref{Figure:wiggles} is the striking correspondence between 8 (possibly 9) of the condensations (vertical dashed lines; Fig.~\ref{Figure:wiggles}) and the extremes in amplitude in the velocity field. The probability of finding a column density peak at the velocity extrema can be approximated as $p=~2\langle R_{\rm eq}\rangle/(\lambda_{\rm vel, i}/2)\sim~0.36$ (where $\langle R_{\rm eq}\rangle$ is the mean equivalent radius; Table~\ref{Table:clouds}). The binomial probability, $P$, of finding at least 8 column density peaks at the velocity extrema is therefore $P\sim0.03$, implying that this correspondence is statistically significant at the 97 per cent confidence level.

\section{Seeding the Galactic Centre gas stream}

The correspondence between the column density peaks and the extremes in the velocity field implies that gas is accumulating at locations where the local velocity differential is minimal. These `traffic jams' mean that gas must locally accumulate, potentially allowing gravitational instabilities to take over and play a key role in the formation of the molecular clouds on the GC gas stream. We investigate this hypothesis by considering the stability criteria for gas clouds in galactic discs, described by \citet{toomre_1964}. On size scales smaller than the Jeans length,
\begin{equation}
\lambda_{\rm J}\approx\frac{\sigma^{2}_{v}}{G\Sigma}
\label{Eq:jeans}
\end{equation}
where $\sigma_{v}$ is the velocity dispersion of the gas (including contributions to the gas pressure from both thermal and non-thermal motions), $\Sigma$ is the gas surface density, and $G$ is the gravitational constant, molecular gas is stabilised against gravitational collapse by internal pressure gradients. Similarly, in a rotating galactic disc, on size scales greater than the Toomre length,
\begin{equation}
\lambda_{\rm T}\approx4\pi^{2}\frac{G\Sigma}{\kappa^{2}}
\label{Eq:toomre}
\end{equation}
where $\kappa$ refers to the epicyclic frequency, molecular gas is stabilised against gravitational collapse by rotation and shear. This defines a size scale, $\lambda_{\rm J}<\lambda<\lambda_{\rm T}$ over which gravitational instabilities are able to grow, leading to the formation of gravitationally unstable molecular clouds. 

At large galactocentric radii ($R_{\rm GC}\gtrsim150$\,pc), but still within the CMZ, the rotation curve of the Galaxy is near-flat \citep{launhardt_2002, bhattacharjee_2014}. In this regime, the shear and velocity dispersion of the gas is high, and surface density low in comparison to the GC gas stream. The combination of these factors leads to $\lambda_{\rm J}>\lambda_{\rm T}$, and stability against gravitational collapse (i.e.~Toomre $Q\gg1$). Observationally, this is supported by low levels of star formation activity within giant complexes such as Bania's clump 2 and the 1\fdg3 cloud complex (e.g. \citealp{bally_2010}). However, at a galactocentric radius of $R_{\rm GC}\sim120$\,pc the rotation curve transitions from near-flat to near solid body \citep{krumholz_2015}. Observationally, this radius encompasses the GC gas stream, where the surface density is high and velocity dispersion is low by comparison (although still of the order $\sim10$\,\kms; \citealp{henshaw_2016}). These factors, coupled with the reduction in shear, lead to $\lambda_{\rm J}<\lambda_{\rm T}$, potentially allowing the gas to become gravitationally unstable.

In the models of \citet{kruijssen_2014b}, \citet{krumholz_2015} and \citet{krumholz_2016}, gas at $R_{\rm GC} > 120~{\rm pc}$ is stable against collapse due to acoustic instabilities induced by the Galactic bar. These instabilities drive both turbulence and angular momentum transport in the CMZ gas, causing the gas to flow inwards and simultaneously increase in velocity dispersion. As the rotation curve of the Galaxy transitions from near-flat to near solid body, the reduction in shear suppresses transport and turbulent driving, leading to an accumulation of material which can become gravitationally unstable, akin, argue the authors, to the 100\,pc gas stream. 

Using Equations~\ref{Eq:jeans} and \ref{Eq:toomre} we can estimate the relevant instability lengths for the gas stream. Assuming $\sigma_{v}\sim5$\,\kms, the mean value of the velocity dispersion measured via HFS fitting at the peak location of each of the leaves (after correcting for the contribution to the velocity dispersion from ordered velocity gradients; panel B, Fig.~\ref{Figure:wiggles}), $\Sigma\sim1000$\,\solar\,pc$^{-2}$, an order of magnitude estimate for the surface density throughout this portion of the gas stream, and $\kappa\sim3.2$\,Myr$^{-1}$ \citep{launhardt_2002}, we find $\lambda_{\rm J}\sim6$\,pc and $\lambda_{\rm T}\sim17$\,pc, respectively. 

Computing the minimum projected nearest-neighbour separation between peaks of each of the molecular clouds identified in panel A of Fig.~\ref{Figure:wiggles} we find $\lambda_{\rm sep}\sim7$\,pc, and after correcting for inclination we find $\lambda_{\rm sep, i}\sim8$\,pc. Since the smallest gas condensations are separated by a Jeans length ($\lambda_{\rm sep,i}\sim\lambda_{\rm J}$) and the largest corrugation scale matches the Toomre length ($\lambda_{\rm vel, i}\sim\lambda_{\rm T}$), structure grows over the same range of size-scales over which the gas can be gravitationally unstable. Gravitational instabilities are therefore likely driving the formation of structure within the GC gas stream. 

In the context of the above scenario, the observed corrugations may be similar, albeit on larger scales, to patterns reported within individual molecular clouds within the Galactic disc which have been interpreted as mass accretion (\citealp{hacar_2011, henshaw_2014}; Arzoumanian et al. in preparation). Close correspondence between the virial velocity of the condensations, $v_{\rm vir}=~\sqrt{G\langle M \rangle/\langle R_{\rm eq} \rangle} \sim~5$\,\kms, and the amplitude of the corrugations ($3.7\,\pm0.1$\,\kms) supports this hypothesis. Such an interpretation, however, is intimately linked to the 3-D geometry of the system (\citealp{henshaw_2014, gritschneder_2016}). Indeed, if the corrugation in the GC gas stream signifies gas accretion, the correspondence between the extremes in the density and velocity fields would necessitate that the stream itself exhibits a physical undulation. Such a configuration is qualitatively supported by observations of external galaxies, in which physical undulations and corresponding velocity corrugations have been reported that are interpreted as the result of large-scale gravitational instabilities (e.g. IC~2233; \citealp{matthews_2008b}). 

It remains unclear whether the velocity pattern represents the cause or the effect of gravitational instabilities in the gas stream. In the former scenario (cause), the observed corrugation in the velocity field generates `traffic jams' in the gas flow that seed the convergence towards the centres of mass of the Toomre-scale condensations. If this is indeed true, then galactic dynamics (i.e. shear) set the oscillation length, below which fragmentation occurs on the local Jeans length. Smaller clouds cannot form due to stabilisation by internal pressure. The latter scenario (effect) represents the reverse causality. Here, the similarity of the Jeans and Toomre lengths (modulo a factor of $\sim2$) allows gravitational instability on a fixed length-scale, resulting in the local convergence of gas flows of which the velocity corrugations are merely an observational tracer rather than the cause. In spite of this ambiguity, the key result of the above calculation is not the absolute values of the derived length scales themselves, but the fact that they are similar and of the order 10 pc. This is very similar to the observed wavelengths of the velocity fluctuations and the separation lengths of the condensations. {\it This implies that the gas stream is marginally gravitationally unstable, and that gravitational collapse is only just setting in.}

Qualitative comparison between the condensations and the dust ridge molecular clouds, to which the upstream gas connects contiguously \citep{henshaw_2016}, supports this conclusion. The spacing of the dust ridge clouds is similar to that of the condensations, of the order $\sim12$\,pc. This leads us to speculate that the dust ridge molecular clouds may also have formed as a result of gravitational instabilities in the gas stream. However, the condensations associated with the corrugations are systematically less massive than those on the dust ridge ($\sim10^{3}$-$10^{4}$\,\solar \ versus $\sim10^{4}$-$10^{5}$\,\solar; see Table~\ref{Table:clouds} and \citealp{walker_2015}), even though they have similar sizes ($R_{\rm eq}\sim2-3$\,pc). As a consequence of this result, the typical surface density contrast between peak and stream is only of the order $\lesssim10$, much smaller than for, e.g., G0.253+0.016, which displays a density contrast between peak and stream of several tens \citep{longmore_2012}. Coupling this observation with the lack of star formation activity reported towards this portion of the gas stream, leads us to infer that our observed condensations are in an early evolutionary phase and may reach the masses and densities of the dust ridge clouds by accreting more gas over the next free-fall time ($t_{\rm ff,eq}=0.37$--$0.63$~Myr, see Table~\ref{Table:clouds}). In the \citet{kruijssen_2015} orbital model, the velocity corrugations are 0.3--0.8 Myr upstream from the dust ridge cloud G0.253+0.016, implying that they will undergo $\sim$ one more free-fall time before reaching the current position of the dust ridge.

As demonstrated by both \citet{kruijssen_2015} and \citet{henshaw_2016}, the distribution of material throughout the GC gas stream is intrinsically inhomogeneous. This is reflected in variations in the velocity pattern, possibly due to local variations in the surface density, evident in Fig.~\ref{Figure:wiggles} (panel C). The initial conditions for individual clouds, for instance those in the dust ridge, will therefore depend intimately on the historical conditions of the stream. Future work, using empirically-derived physical and dynamical constraints (Henshaw et al. in preparation) in conjunction with hydrodynamical simulations investigating the tidal interaction between molecular clouds and the Galactic gravitational potential during pericentre passage (Kruijssen et al. in preparation), will investigate the prospect that the identified condensations are analogous to the progenitors of the dust-ridge molecular clouds. If confirmed, the velocity corrugations and corresponding condensations may represent observational signatures of the initial phases of star and cluster formation, pinpointing the beginning of an observable evolutionary time sequence of star formation.

\section*{Acknowledgements}

We would like to thank the referee, Erik Rosolowsky, for the constructive report which has helped to improve the paper. We thank Cara Battersby for providing the \emph{Herschel}-derived column density map used throughout this work. This research has benefitted from the {\sc astropy} \citep{astropy_2013}, {\sc matplotlib} \citep{hunter_2007}, and {\sc astrodendro} (\url{www.dendrograms.org}) software packages. JMDK gratefully acknowledges financial support in the form of a Gliese Fellowship and an Emmy Noether Research Group from the Deutsche Forschungsgemeinschaft (DFG), grant number KR4801/1-1.

%%%%%%%%%%%%%%%%%%%%%%%%%%%%%%%%%%%%%%%%%%%%%%%%%%



\bibliographystyle{mnras}
\bibliography{References/references} % if your bibtex file is called example.bib

%%%%%%%%%%%%%%%%%%%%%%%%%%%%%%%%%%%%%%%%%%%%%%%%%%

%%%%%%%%%%%%%%%%% APPENDICES %%%%%%%%%%%%%%%%%%%%%

\appendix

%%%%%%%%%%%%%%%%%%%%%%%%%%%%%%%%%%%%%%%%%%%%%%%%%%

\bsp	% typesetting comment
\label{lastpage}
\end{document}